\documentstyle[prl,aps,12pt,epsf]{revtex}
\def\inseps#1#2{\def\epsfsize##1##2{#2##1} \centerline{\epsfbox{#1}}}
\tightenlines

\def\be{\begin{equation}}
\def\th0{\theta_0}

\def\bea{\begin{eqnarray}}
\def\ee{\end{equation}} 
\def\eea{\end{eqnarray}} 

\begin{document}
\title{Wave propagation in a chiral fluid: an undergraduate study}

\author{Thomas Garel}

\address{Service de Physique Th\'eorique, CEA/DSM/SPhT \\
Unit\'e de recherche associ\'ee au CNRS \\
CEA/Saclay,
91191 Gif-sur-Yvette, Cedex, France.}


\maketitle

\vskip 50mm

\noindent\mbox{Submitted for publication to: ``European Journal of Physics''} \hfill 
\mbox{Saclay, T03/046}\\ \noindent \mbox{ }\\ 

\bigskip


\newpage

{\bf Abstract}

We study the propagation of electromagnetic waves in a chiral fluid,
where the molecules are described by a simplified version of the Kuhn
coupled oscillator model. The eigenmodes of 
Maxwell's equations are circularly polarized waves. The application of
a static magnetic field further leads to a magnetochiral term in the
index of refraction of the fluid, which is independent of the wave
polarization. A similar result holds when absorption is taken into
account. Interference experiments and photochemical reactions have
recently demonstrated the existence of the magnetochiral term. The
comparison with Faraday rotation in an achiral fluid emphasizes the
different symmetry properties of the two effects.
   

\bigskip

\bigskip

\bigskip

\bigskip

{\bf R\'esum\'e}

On \'etudie la propagation d'ondes \'electromagn\'etiques dans un
fluide chiral, dont les mol\'ecules sont d\'ecrites \`a l'aide d'une
version simplifi\'ee du mod\`ele d'oscillateurs coupl\'es de
Kuhn. Les modes propres des \'equations de Maxwell sont 
des ondes polaris\'ees circulairement. L'application d'un champ
magn\'etique statique entra\^\i ne l'existence d'un terme
magn\'etochiral dans l'indice de r\'efraction du fluide,
ind\'ependamment de la polarisation de l'onde. Un r\'esultat semblable
s'applique au cas de l'absorption. Le terme magn\'etochiral a \'et\'e
r\'ecemment mis en \'evidence dans des exp\'eriences d'interf\'erences
et dans des r\'eactions photochimiques. La comparaison \`a la rotation 
Faraday dans un fluide achiral souligne les diff\'erences de
sym\'etrie des deux cas.

\bigskip

\bigskip

\bigskip

\bigskip

\newpage

\bigskip

\newpage
\section{Introduction}
\label{Intro}
In simple fluids, the propagation of an electromagnetic wave is
usually treated as follows. Sticking to a classical description,
one considers that the bound electrons of the fluid molecules are
displaced from their equilibrium positions by the Lorentz force of the
wavefield. This (time dependent) displacement 
or induced electric dipole can be modelled as a current density ${\rm {\vec j}}
(\vec r,t)$ which is related to the electromagnetic field ${\rm {\vec
E}} (\vec r,t)$ and ${\rm {\vec B}} (\vec r,t)$ through Newton's force 
equation (Complications such as local fields effects will not be
considered in this paper). Neglecting the magnetic contribution (the
Bohr model  of the hydrogen atom leads to an estimate $\vert {F_{magn} \over
F_{elec}}\vert \sim \vert{\vec v \times \vec {\rm B} \over \vec {\rm
E}}\vert \sim {v \over c} < 10^{-2}$), and using Maxwell's equations
then leads to textbook expressions for the index of refraction of the fluid 
\cite{feyn}

For chiral molecules, one has to consider the displacement of the bound
electrons on a well defined geometric structure. In other words, one has
to take into account the variation of the electromagnetic field down
to molecular distances. The induced electric dipole moment 
(or current density) depends now on the field and on its spatial
derivatives. This may be the reason why the propagation of
electromagnetic waves in a chiral fluid is not frequently treated at
an undergraduate level \cite{LL}.  

On the other hand, students may encounter phenomenological models which
show that Maxwell's equations in an isotropic chiral fluid have
circularly polarized waves as eigenmodes \cite{US
problems,Sil_Soh,Nie_Pal}. This in turn suggests  
a comparison with the Faraday effect \cite{Van}, which deals with the
propagation of waves in an simple (achiral) fluid in the presence of a
magnetic field. The comparison of these two situations raises
questions about parity and time reversal
transformations in a ``non particle physics'' context
\cite{Rin_Cal,Barron3,Siva,Bar_Zel}. 
These symmetry considerations are not confined to theory: recent
experiments \cite{Rik_Rau1,Kle_Wag,Rik_Rau2} have shown that 
the interplay of chirality and magnetism have remarkable consequences,
that have been observed in interference experiments and
in photochemical reactions. Roughly speaking, the application of a
magnetic field on a chiral fluid leads to a change in the index of
refraction, and this new ``magnetochiral'' term is independent of
the state of polarization of the wave. The order of magnitude of this
term is rather small (see below), but its  possible implications for
the origin of terrestrial homochirality in biomolecules (DNA,
proteins,...) are rather interesting  \cite{Wag_Mei}.

The present paper does not claim to be original. Its merit is perhaps
to bring to the attention of students some topics which are somehow
scattered through the litterature. Furthermore, experiments which are both
``hot'' and  accessible to undergraduates are not so frequent. 

To illustrate these considerations in a self contained way, section
\ref{convent} will present polarization conventions and notations, as
well as some definitions pertaining to chirality. Section \ref{faraday}
presents a sketchy  
derivation of the Faraday effect in a simple (achiral) fluid. In the
framework of the elastically bound electron model, this model yields
reasonable orders of magnitude, and enables one to study various
aspects of the Faraday effet (dependence on the direction of
propagation of light with respect to the magnetic field, reflection on 
a conducting mirror). The introduction of a damping term in the
electron's equation of motion can then model absorption phenomena.

In section \ref{chiral}, we consider a model of chiral molecules, which
consists of two coupled anisotropic 
oscillators \cite{Kuhn,Cal_Eyr}. By further restricting the accessible 
orientations (albeit in an isotropic way), we will derive to lowest
order the influence of chirality on Maxwell's equations and show
explicitly that the eigenmodes are circularly polarized waves.
The differences with the Faraday effect will be pointed out, in
particular from the symmetry point of view.

In section \ref{magnetic}, we consider the effect of a magnetic
field on a chiral fluid, along the lines of Larmor's theorem. This
non-rigorous point of view \cite{Bar_Zel} suggests the existence of a
new (symmetry allowed) term in the index of refraction, called the
magnetochiral term. Its order of magnitude is obtained through the
model of section \ref{chiral}, and we discuss a recent interference
experiment where this term is involved. Finally, when absorption is
taken into account, the magnetochiral contribution leads to remarkable
results in photochemical reactions.

\section{Definitions and notations}
\label{convent}
Two molecules are superposable if one can bring them into coincidence
by using only translations and rotations. A molecule (${\rm L}$) is
chiral if it is non-superposable onto its 
mirror image (${\rm D}$). The (${\rm L}$) and (${\rm D}$) forms of a
molecule are called enantiomers of this molecule. A fluid made of
equal amounts of both enantiomers is called racemic.
A fluid made only of (${\rm L}$) (resp. (${\rm D}$) ) molecules will
be called a (${\rm L}$)-fluid (resp. a (${\rm D}$)-fluid) in this paper.

In agreement with many authors
\cite{Cal_Eyr,Mason,Barronn,Lowry}, we consider a wave to be right
(resp. left) 
circularly polarized wave if an observer looking towards the source sees
at a fixed point in space, the tip of the electric field turn
clockwise (resp. anticlockwise) with time.

Physical properties pertaining to a right (resp. left) circularly
polarized wave will be denoted by the subscript {\bf +} (resp. {\bf -}).
A subscript ${\rm L}$ (resp. ${\rm D}$) will be further added if one deals
with a (${\rm L}$)-fluid (resp. (${\rm D}$)-fluid).
So $n_{+}^{\rm L}(\omega)$  denotes the index of refraction of a right
circularly polarized wave of frequency $\omega$ in the $({\rm L})$-fluid.

\section{The Faraday effect}
\label{faraday}
\subsection{The elastically bound electron model}

We consider a monochromatic plane wave, of frequency $\omega$,
propagating in a simple (achiral) fluid. The direction of propagation
is the $z>0$ direction. We 
model the atoms as hydrogen atoms, where the bound electrons are
submitted to an elastic force $\vec 
f=-m_e\omega_0^2 \vec r$, where $m_e$ is the electron mass, $\omega_0$ 
a typical electronic frequency ($\omega_0 \sim 10^{15}$ Hz), and $\vec
r$ is measured from the equilibrium position of the
electron. Neglecting the magnetic force, we get
\be
\label{force}
m_e \ddot{\vec r}=-m_e\omega_0^2 \vec r-e\vec {\rm E}
\ee
If $N$ is the particle density, the induced electric dipole moment per unit
volume then reads
\be 
\label{pol1}
{\rm {\vec P}}=-Ne\vec r=+{Ne^2\vec {\rm E} \over {m_e(\omega_0^2-\omega^2)}}
\ee
leading to a polarisation current density ${\rm {\vec j}}= {\partial {\rm
{\vec P}} \over
\partial t}$. Plugging these results in Maxwell's equations
\cite{transverse}, and denoting by $c$ the speed of light in vacuo yields 
\be
\Delta {\rm {\vec E}}={1 \over c^2}(1+{Ne^2 \over
m_e\varepsilon_0(\omega_0^2 -\omega^2)}) \ {\partial^2{\rm {\vec E}}
\over \partial t^2}
\ee

Looking for a plane wave solution $\vec {\rm E}={\rm Re}(\vec {\rm E}_0 
e^{j(kz-\omega t)})$, and introducing the index of refraction
$n(\omega)$ through $k={\omega \over {c/n(\omega)}}$ , we get 

\be
n^2(\omega)=1+{Ne^2 \over
m_e\varepsilon_0(\omega_0^2 -\omega^2)}
\ee

\subsection{Effect of a magnetic field and circular polarizations}

We specifically consider the propagation of a right circularly
polarized wave (${\rm {\vec E_+}}={\rm {E}}_0 ({\rm
cos}(kz-\omega t), {\rm sin}(kz-\omega t), 0)$) in the presence of a
static magnetic field $\vec B_0=B_0 \vec e_z$. The equation of motion 
\be
\label{force2}
m_e \ddot{\vec r}=-m_e\omega_0^2 \vec r-e({\rm {\vec E_+}}+\dot{\vec r} 
\times {\vec B_0})
\ee
shows that the Lorentz force due to $B_0$ is antiparallel to the
elastic force. Following the same steps as before gives 
\be 
\label{nplus} 
n_{+}^2(\omega,B_0)=1+{Ne^2
\over m_e\varepsilon_0(\omega_0^2-\omega^2 -{eB_0 \over m_e}\omega)}
\ee

For a left circularly polarized wave (${\rm {\vec E_{-}}}={\rm {E}}_0
({\rm cos}(kz-\omega t), -{\rm sin}(kz-\omega t), 0)$), we find
\be
\label{nmoins} 
n_{-}^2(\omega,B_0)=1+{Ne^2
\over m_e\varepsilon_0(\omega_0^2-\omega^2 +{eB_0 \over m_e}\omega)}
\ee
since in this case, the Lorentz force due to $B_0$ is parallel to the
elastic force.

\subsection{Faraday rotation}

Let us now consider a linearly polarized wave propagating in a confined
fluid ($0<z<l$), in the presence of the magnetic field $\vec B_0=B_0
\vec e_z$.
Since a linearly polarized wave can be decomposed into two circularly
polarized waves $(+)$ and $(-)$ of equal amplitude but with different
phase velocities ($v_{\pm}=c/n_{\pm}(\omega,B_0)$), elementary
calculations show that the direction of vibration of the electric
field rotates between $z=0$ and $z=l$ by an amount $\alpha$, with 

\be
\alpha={\omega l \over 2c} \ \Delta n(\omega,B_0) 
\ee
where $\Delta n(\omega, B_0)=n_{+}(\omega,B_0)-n_{-}(\omega,B_0)$.
This rotation is counterclockwise for an observer looking towards the
source since $v_{+} <v_{-}$ as shown by equations
(\ref{nplus},\ref{nmoins}). (Note that this simple model applies only
to diamagnetic 
materials; paramagnetic materials require a more sophisticated
treatment).

Equations (\ref{nplus},\ref{nmoins}) also show that, as long as
one deals with frequencies $\omega \gg \omega_L= {eB_0 \over
2m_e}$, one has
\be 
\label{Larmor}
n_{\pm}(\omega,B_0)\simeq n(\omega \pm \omega_L)
\ee
up to ${\rm O}({\omega_L \over \omega})^2$ terms.
Since $\omega \sim \omega_0 \sim 10^{15}$ Hz and even for $B_0=10$ T,
we only have $\omega_L \sim 10^{12}$ Hz, we will consider that Larmor
theorem, as expressed by equation (\ref{Larmor}), holds. This allows
us to rewrite the magnetic birefringence as

\be
\label{Larmor2}
\Delta n(\omega, B_0)\simeq {eB_0 \over {m_e}} {dn(\omega) \over
d\omega}
\ee
Typical orders of magnitude for a liquid such as ${\rm CS}_2$ are
$\Delta n(\omega, B_0) \sim 10^{-5}$ for $B_0=1$ T. The observation of
the Faraday birefringence in an interference experiment is relatively easy
since the Rayleigh and Michelson interferometers may detect index
variations down to $10^{-8}$ \cite{Bor_Wol}. 

\medskip 

The above expressions pertain to a wave propagating parallel to the
magnetic field $\vec B_0$. For propagation in direction $\vec u$ (with
$\vec u={\vec k \over k}$), equations (\ref{Larmor},\ref{Larmor2}) read

\be
\label{Larmor3}
n_{\pm}(\omega,\vec B_0)\simeq n(\omega \pm {e\vec B_0 \cdot \vec u
\over 2m_e})
\ee

and
\be
\label{Larmor4}
\Delta n(\omega, B_0)\simeq {e
({\vec  {B_0}} \cdot \vec {u}) \over {m_e}} {dn(\omega) \over
d\omega}
\ee

as can be checked for a wave propagating in the $z<0$ direction ($\vec
B_0 \cdot \vec u <0$).

\bigskip 

The effect of placing a perfectly conducting mirror at $z=l$ in the
original experiment (where the incident wave has $\vec u=\vec e_z$)
can be analyzed from 
different points of view. The simplest one 
is probably described in \cite{Siva}: for an observer receiving the
reflected wave, the incident wave from $z=0$ to $z=l$ is equivalent,
in the (symmetric w.r.t the mirror) image space, to a wave
propagating in the $z<0$ direction in the same magnetic field $\vec
B_0=B_0 \vec e_z$
(this is a clear illustration of the axial character
of the magnetic field $\vec B_0$, since $\vec B_0$ is perpendicular to
the mirror). Taking into account the reflected wave, shows 
that Faraday's rotation is doubled for this observer (the reflection
on the mirror is irrelevant since it changes the sense, but not the
direction of the electric field). Typical orders 
of magnitude for a liquid such as ${\rm CS}_2$ are $\alpha \simeq
0.2-0.3$ rd, for $l=10^{-2}$ m and $\Delta n(\omega, B_0) \sim 10^{-5}$.

 \bigskip

Finally, the inclusion of damping in equations
(\ref{force},\ref{force2}) leads to an absorption of the wave by the
fluid. Denoting by $n_2$ the imaginary part of the index of
refraction, we clearly have $n_{2+}(\omega ,B_0) \ne n_{2-}(\omega
,B_0)$. This magnetic dichroism implies an elliptical polarization
(together with a Faraday rotation of the major axis of the ellipse) in
the above experiments. Typical orders of magnitude are ${\Delta
n_2(\omega,B_0) \over n_2(\omega)} \sim 10^{-4}-10^{-5}$ for $B_0=1$
T.

\section{Propagation in a chiral fluid}
\label{chiral}

\subsection{The simplified Kuhn model}
\label{Kuhn}
As previously mentionned, to take chirality into account requires a
rather detailed geometric description of the fluid molecules. We
consider here the simplest model of a chiral molecule
\cite{Kuhn,Cal_Eyr}, which consists of two coupled oscillators.

To fix notations, we consider a fixed trihedron ($Oxyz$), with unit
vectors ($\vec e_x, \vec e_y, \vec e_z$), and a $({\rm
L})$-fluid with $N$ molecules per unit volume. Each molecule has two
electrons, whose equilibrium positions are called $\vec R_1^{0}$ and
$\vec R_2^{0}$. We denote by $d=\vert \vec R_2^{0}-\vec R_1^{0}
\vert$ the size of the molecule. We also define $\vec R_{12}^{0}=\vec
R_2^{0}-\vec R_1^{0}=d  \ \vec {b_{12}^0}$ and $\vec R_0={\vec R_1^0 +\vec
R_2^0 \over 2}$. 
 
Due to their interaction with the propagating wave, the electrons are
displaced from $\vec R_1^{0}$ (resp. $\vec R_2^{0}$) by an amount $\vec
r_1$ (resp. $\vec r_2$).  

For a fixed orientation of the molecule (i.e. for a fixed $\vec
{b_{12}^0}$) , the displacements of the Kuhn model are both one
dimensional and coupled:

(i) the unit vector in the direction of
$\vec r_1$ (resp. $\vec r_2$) is denoted by $\vec b_1$ (resp. $\vec
b_2$). The chirality stems from the fact that the unit vectors
$\vec b_1$, $\vec b_2$ et $\vec b_{12}^{0}$ are such that
\be
\label{chiral1}
\chi=\vec {b_{12}^0} \cdot (\vec b_1 \times \vec b_2)=-1
\ee

(ii) the potential energy of the two electrons is written as 

\be
E_p={1 \over 2}m_e \omega_0^2 ({\vec r_1}^{2}+{\vec r_2}^{2})+m_e
\Omega_{12}^{2} (\vec b_1 \cdot \vec r_1) (\vec b_2 \cdot \vec r_2) 
\ee
where $\omega_0$ and $\Omega_{12}$ are electronic frequencies.

\medskip

Compared to the original Kuhn model, we further add two restrictions
\cite{Cal_Eyr}:

(iii) With respect to the fixed $(Oxyz)$ trihedron, each molecule of the
fluid can only adopt the six orientations (${\rm A_i}, \
{\rm i}=1,2,...,6$) shown in Figure 1.

The geometrical parameters of (${\rm
A_i}$) are denoted with an index $({\rm i})$: we thus have $\vec
b_{1}^{({\rm i})}, \vec b_{2}^{({\rm i})}, \vec r_{1}^{({\rm i})}, \vec
r_{2}^{({\rm i})},.....$. The chirality is clearly the same for all
orientations 
\be
\chi^{({\rm i})}=\vec {b_{12}^{0({\rm i})}} \cdot (\vec b_1^{({\rm
i})} \times \vec b_2^{({\rm i})})=\chi=-1. 
\ee

(iv) We assume that the orientations  (${\rm A_i}$) are
equiprobable. 

\medskip

These restrictions preserve the isotropy of the fluid,
and make the calculations easier (more complete calculations can be
found in ref \cite{Cal_Eyr}).

\subsection{Equations of motion}
\label{motion}

Neglecting the magnetic contribution to the Lorentz force, we get

\be
\label{mouv1}
m_e{\ddot {\vec r_1}}=-m_e\omega_0^2 \vec r_1-m_e\Omega_{12}^2\vec b_1 
(\vec b_2 \cdot \vec r_2)-e{\rm {\vec E}}(\vec r_1)
\ee

and

\be
\label{mouv2}
m_e{\ddot {\vec r_2}}=-m_e\omega_0^2 \vec r_2-m_e\Omega_{12}^2\vec b_2 
(\vec b_1 \cdot \vec r_1)-e{\rm {\vec E}}(\vec r_2)
\ee

Projecting on $\vec b_1$ and $\vec b_2$, and looking for forced
solutions, we obtain for orientation (${\rm A_i}$)
\be
{\vec r_1 \choose \vec r_2}^{({\rm i})}={\rm M}^{-1} \ { {-e{\rm {\vec
E}}(\vec r_1) \cdot \vec b_1 \over m_e} \choose {-e{\rm {\vec
E}}(\vec r_2) \cdot \vec b_2 \over m_e}}^{({\rm i})}
\ee

where the orientation independent matrix $\rm M$ is given by
$$
{\rm M}=\pmatrix{
\omega_0^2-\omega^2&\Omega_{12}^2\cr
\Omega_{12}^2&\omega_0^2-\omega^2\cr
}$$

and ${\rm M}^{-1}$ is its inverse.

Let us
consider for the time being a linearly polarized wave, with $\vec {\rm
E}={\rm Re}({\rm E}(z) \ e^{-j\omega t}) \vec e_x$.
The resolution of the equations for each (equiprobable) orientation 
yields the induced electric dipole moment ${\rm {\vec P}}$ of the fluid. It
will turn out that ${\rm {\vec P}}$  has components both parallel
(${\rm {\vec P}}_{//}$)
and perpendicular (${\rm {\vec
P}}_{\perp}$) to 
the field $\vec {\rm E}$. It is easily seen that orientations
$({\rm A_5},{\rm A_6})$ do not contribute to ${\rm {\vec P}}$, and that
orientations $({\rm A_3},{\rm A_4})$ only contribute to ${\rm {\vec
P}}_{//}$. On 
the other hand, orientations $({\rm A_1},{\rm A_2})$ contribute to both 
components. Elementary algebra show that
\be
\label{perp}
{\rm {\vec P}}_{\perp}={Ne^2 \over 6m_e} {\Omega_{12}^2 \over
{(\omega_0^2-\omega^2)^{2}-\Omega_{12}^{4}}} ({\rm E}(z_{1}^{(2)})-{\rm
E}(z_{2}^{(1)}) \vec e_y
\ee

Note that the perpendicular dipole component ${\rm {\vec
P}}_{\perp}$ does not vanish because of the finiteness of $d$ (see
Figure 1). To lowest order in $d$ (or more appropriately in ${d \over
\lambda}$), one gets 
\be 
{\rm {\vec P}}_{\perp}=-{Ne^2 \over 6m_e} {\Omega_{12}^2 \over
{(\omega_0^2-\omega^2)^{2}-\Omega_{12}^{4}}} \ (d \ {\partial{\rm E}(z) \over 
\partial z}) \ \vec e_y
\ee
where the gradient term is calculated at $\vec R_0$.

Gathering all contributions to ${\rm {\vec P}}$ and considering a
general state of polarisation of the wave, we get
\be 
\label{pol}
\vec {\rm P}=\alpha(\omega) \ \vec {\rm E} +\gamma^{\rm L}(\omega) \ {\rm curl}
\ \vec {\rm E } 
\ee
where the electric field and its derivatives are calculated at $\vec
R_0$. In equation (\ref{pol}), one has

\be
\label{alpha}
\alpha(\omega)={2Ne^{2} \over 3m_e}{(\omega_0^2-\omega^2) \over
{(\omega_0^2-\omega^2)^{2}-\Omega_{12}^{4}}}
\ee
and
\be
\label{gamma}
\gamma^{\rm L}(\omega)={Ne^{2} \over 6m_e}{\Omega_{12}^2 (\vec {R_{12}^0}
\cdot (\vec b_1 \times \vec b_2)) \over
{(\omega_0^2-\omega^2)^{2}-\Omega_{12}^{4}}}
\ee
where we have used equation (\ref{chiral1}).
Equation (\ref{gamma}) explicitly shows that $\gamma^{\rm L}(\omega)$
is a pseudo (or axial) scalar.
This feature of $\gamma^{\rm L}(\omega)$ is indeed required by equation
(\ref{pol}), since ${\rm {\vec P}}$ is a true (or polar) vector and
${\rm curl} \ \vec {\rm E }$ a pseudo (or axial) vector \cite{moma}.

\subsection{Rotatory power}
\label{rotatory}
Using equation (\ref{pol}) together with  $\vec {\rm j}= {\partial
{\rm {\vec P}} \over \partial t}$, one finds \cite{transverse2}
\be
\label{propchir}
\Delta {\rm {\vec E}}={1 \over c^2}{\partial^2 \over \partial t^2}
\left( (1+{\alpha(\omega) \over \varepsilon_0}){\rm {\vec E}}
+{\gamma^{\rm L}(\omega) \over \varepsilon_0} {\rm curl} \ \vec {\rm E} \right)
\ee

We are now back to the traditional study of Maxwell's equations in a
chiral fluid \cite{US problems,Sil_Soh,Nie_Pal}. For circularly
polarized waves in the long
wavelength approximation (${d \over \lambda} \ll 1$), we get from
equation (\ref{propchir})
\be
\label{chirali}
n_{\pm}^{\rm L}(\omega) \simeq n(\omega)\mp {\gamma^{\rm L}(\omega) \over
2\varepsilon_0} k_0
\ee
with $n^2(\omega)=1+{\alpha \over
\varepsilon_0}$ and $k_0={\omega \over c}$.

A rough order of magnitude for the natural birefringence ($\Delta
n^{\rm L}(\omega)=n_{+}^{\rm L}(\omega)-n_{-}^{\rm L}(\omega)$) of a
chiral liquid can be obtained from $d=10$ \AA, $N \sim 5 \ 10^{26}$ m$^{-3}$,
$\omega=\Omega_{12}={\omega_0 \over 2}= \pi \ 10^{15}$ Hz, leading to 
$\Delta n^{\rm L}(\omega)\simeq 3 \cdot 10^{-5}$. 

Several points can easily be checked on equations
(\ref{propchir},\ref{chirali}) 

\medskip

(i) the index of refraction is the same for a (e.g. right) circularly
polarized wave propagating in the (${\rm L}$)-fluid, in the $z>0$ and
$z<0$ directions.

\medskip

(ii) for the (${\rm D}$)-fluid, one has $n_{\pm}^{\rm
D}(\omega)=n_{\mp}^{\rm L}(\omega)$, since $\gamma^{\rm
L}(\omega)=-\gamma^{\rm D}(\omega)$.

\medskip

(iii) if a linearly polarized wave propagates in a $({\rm
L})$-fluid, confined between $z=0$ and $z=l$, the direction of
vibration of the electric field will rotate by an angle $\beta$ given
by
\be
\beta={k_{0}l \over 2}\Delta n^{\rm L}(\omega)={\vert\gamma^{\rm
L}(\omega)\vert k_{0}^2 l \over 2\varepsilon_0} 
\ee
In our model of a $({\rm L})$-fluid, the rotation is
counterclockwise, and a typical order of magnitude is $\beta
\sim 1.5$ rd for $l \sim 10^{-2}$ m.

\medskip

(iv) if one places a perfectly conducting plane
at $z=l$, we again follow \cite{Siva}:  for an observer receiving the
reflected wave, the incident wave is equivalent to a wave propagating
in the (symmmetric w.r.t the mirror) image space in the $z<0$
direction. In the image space, the image of the $({\rm L})$-fluid is
the $({\rm D})$-fluid (again a clear illustration of the pseudo-scalar
character of chirality). The full experiment for this observer
amounts to a propagation in the $z<0$ direction, first through the
$({\rm D})$-fluid (incident wave), and then through the $({\rm
L})$-fluid (reflected wave), leading to a cancellation of the angle
of rotation. Once again, the reflection
on the mirror is irrelevant since it changes the sense, but not the
direction of the electric field. This cancellation contrasts with the
Faraday result and emphasizes the symmetry differences between the two 
cases \cite{LL,Barron3,Siva}
 
\bigskip

Finally, absorption of the wave can be modelled through damping terms 
in equations (\ref{mouv1},\ref{mouv2}). Detailed calculations can be found
in \cite{Cal_Eyr}. We only stress here that the symmmetry properties of the 
real and imaginary parts of the index of refraction are very
similar. In particular, we have $n_{2+}^{\rm L}(\omega) \ne 
n_{2-}^{\rm L}(\omega)$ (natural dichroism) and $n_{2\pm}^{\rm
D}(\omega)=n_{2\mp}^{\rm L}(\omega)$.

\section{Effect of a magnetic field: the magnetochiral effect}
\label{magnetic}

In principle, it is possible to study analytically the effect of a
static magnetic $\vec B_0$ on equations
(\ref{mouv1},\ref{mouv2}). It is much quicker, following
\cite{Bar_Zel} to apply Larmor theorem (\ref{Larmor3}) to equation
(\ref{chirali}) and to write
\be
\label{Larmor5}
n_{\pm}^{\rm L}(\omega,\vec B_0) \sim n_{\pm}^{\rm L}(\omega \pm {e\vec B_0
\cdot \vec u \over 2m_e}) 
\ee
for a wave propagating in direction $\vec u$. Expanding the right hand
side of (\ref{Larmor5}) leads to 
\be
\label{magneto1}
n_{\pm}^{\rm L}(\omega,\vec B_0)= n(\omega) \pm {e({\vec B_0} \cdot {\vec
u}) \over 2m_e} {dn(\omega) \over d\omega} \mp {\gamma^{\rm L}(\omega)k_0 \over
2\varepsilon_0}- {e \over 4m_e\varepsilon_0}{d\gamma^{\rm L}(\omega)
\over d\omega} 
(\vec k_0 \cdot \vec {B_0})
\ee
where $\vec k_0=k_0 \vec u={\omega \over c} \vec u$. Note that the Larmor
frequency shift has not be done on the ``$k_0$'' term of equation
(\ref{chirali}). This seems plausible since the magnetic field $\vec
B_0$ will act only on the coupled oscillators, that is on
$\gamma(\omega)$; the ``$k_0$'' term on the other hand, comes from the
operator ``curl'' in equation (\ref{pol}) which is not affected by the 
dynamics. We refer the reader to reference \cite{Bar_Zel} for further
discussions on this point.

The remarkable prediction of equation (\ref{magneto1}) is that the
last term, called the magnetochiral term, does not depend on the
polarization of the wave. It exists for linearly polarized or non
polarized light, and has several interesting consequences:

(i) magnetochiral birefringence: a (${\rm L}$)-fluid in a magnetic
field and the associate (${\rm D}$)-fluid in the same magnetic
field do not have the same index of refraction for waves propagating
parallel to the field, regardless of the polarization of
the waves (one may also say that a (${\rm L}$)-fluid in a magnetic
field does not have the same index of refraction for waves propagating 
parallel or antiparallel to the field).

(ii) magnetochiral dichroism: the introduction of damping in the Kuhn
model shows that the absorption coefficient of a (${\rm L}$)-fluid in
a magnetic field is not the same as the absorption coefficient of a
(${\rm D}$)-fluid, regardless of the polarization of the wave (one may 
also say that a (${\rm L}$)-fluid in a magnetic field does not absorb
in the same way waves propagating parallel or antiparallel to the field).

(iii) these magnetochiral effects vanish if light propagates
perpendicularly to the field.

(iv) For the mirror experiment described in section \ref{rotatory},
(iv), in the presence of a magnetic field $\vec B_0=B_0 \vec
e_z$, the contributions of the magnetochiral term cancel. The argument 
is the same as for the chiral contributions.

\medskip

From the Kuhn model, one may get an order of magnitude for the
magnetochiral term of equation (\ref{magneto1}): setting $d=10 \AA$,
$N=5 \ 10^{26}$m$^{-3}$,
$\omega=\Omega_{12}={\omega_0 \over 2}=\pi \ 10^{15}$ Hz and $B_0=10$
T, and using equation (\ref{gamma}), we find a contribution to the
index of refraction of 
order $10^{-8}$, quite close to the experimental limits
\cite{Bor_Wol}. One may further argue that this is precisely the order of
magnitude of the terms which have been neglected in Larmor
``theorem''(see equation (\ref{Larmor})). So, one has finally to
turn to experiments, which indeed have demonstrated the existence
of the magnetochiral term(s).

\subsection{Magnetochiral birefringence: an interference experiment}
Imagine a Young double slit experiment with linearly polarized light,
and let $Oz$ be the optical axis of the set-up. After slit $S_1$, one 
adds:
\begin{itemize}
\item{ a vessel $C_1$ containing the $({\rm L})$-fluid in a magnetic
field $\vec B_0=B_0 \vec e_z$  followed by}
\item{another vessel $\overline{C_1}$ containing the associate $({\rm
D})$-fluid in the opposite magnetic field ($-\vec B_0$).}
\end{itemize}
One adds after slit $S_2$:
\begin{itemize}
\item{a vessel $\overline{C_2}$ containing the associate $({\rm
D})$-fluid in a  field $\vec B_0$, followed by}
\item{another vessel ${C_2}$ containing the $({\rm L})$-fluid in the
opposite magnetic field ($-\vec B_0$).}
\end{itemize} 

The vessels are identical. Analyzing this experiment with equation
(\ref{magneto1}) shows that the phase shift of the interference
pattern caused by the enantiomers is entirely due to the magnetochiral
term. All other contributions cancel, whereas the magnetochiral term
is multiplied by four. The experiment has actually been done with a
Michelson interferometer in a slightly modified way \cite{Kle_Wag}.

\subsection{Magnetochiral dichroism: a photochemical reaction}

Discriminating enantiomers is a major problem in chemistry. A
classical experiment is the following (see
\cite{Rau,Inoue} and references therein). One
considers an initially racemic solution, that one illuminates with right
circularly polarized light. The absorption of light gives rise to
chemical reactions (${\rm D} \to {\rm L}$) and (${\rm L} \to  {\rm
D}$) \cite{Thermal}. In a certain range of parameters, these reactions
are first order and we denote by $k_1$ and $k_2$ their respective
rates. Experimentally, one finds $k_1=K I_0 n_{2+}^{{\rm D}}(\omega)$
and $k_2=K I_0 n_{2+}^{{\rm L}}(\omega)$ where $K$ is a constant and
$I_0$ the intensity of the light. The concentrations of the
enantiomers $[{\rm L}]$ and $[{\rm D}]$  are given by  
\be
\label{photo1}
{d[{\rm L}] \over dt}=-{d[{\rm D}] \over dt}=k_1[{\rm D}]-k_2 [{\rm L}]
\ee
At equilibrium, the solution is not racemic anymore, since
one has 
\be
\label{photo2}
y=\left({{[{\rm L}]-[{\rm D}]} \over {[{\rm L}]+[{\rm
D}]}}\right)_{eq}={k_1-k_2 \over k_1+k_2}={n_{2+}^{{\rm
D}}(\omega)-n_{2+}^{{\rm L}}(\omega) 
\over n_{2+}^{{\rm D}}(\omega)+n_{2+}^{{\rm L}}(\omega)}= {n_{2-}^{{\rm
L}}(\omega)-n_{2+}^{{\rm L}}(\omega) 
\over n_{2-}^{{\rm L}}(\omega)+n_{2+}^{{\rm L}}(\omega)}
\ee
so that the least absorbing enantiomer is in excess. 

This experiment has recently been done with
natural (unpolarized) light, in the presence of a magnetic field
parallel to the direction of propagation of light \cite{Rik_Rau2}: the
magnetochiral effect implies that the absorption is not the same 
for the enantiomers, leading again to a non racemic solution at
equilibrium. The experiment of \cite{Rik_Rau2}, achieves a value $y 
\simeq 10^{-4}$, with a magnetic field $B_0=10$ T.

 \section{Conclusion}

We have presented in a simple way old theoretical models
and recent experiments, on ``chirality, light and magnetism''
\cite{Barron4,Barron5,Fer_Del,Ava_Bab}. These topics can
be of interest for undergraduate students, either from a physical or a 
(bio)chemical point of view, the more so since they also show up in
other fields \cite{Opa,Lee_Haa,Lak_Var_Var,Rikk_Fol_Wyd}. At a more 
advanced level, local field effects as well as quantum calculations
can be introduced \cite{Wag_Mei2,Bar_Vrb}.

It is a pleasure to thank J-J. Girerd for discussions.

\newpage

\begin{figure}
\inseps{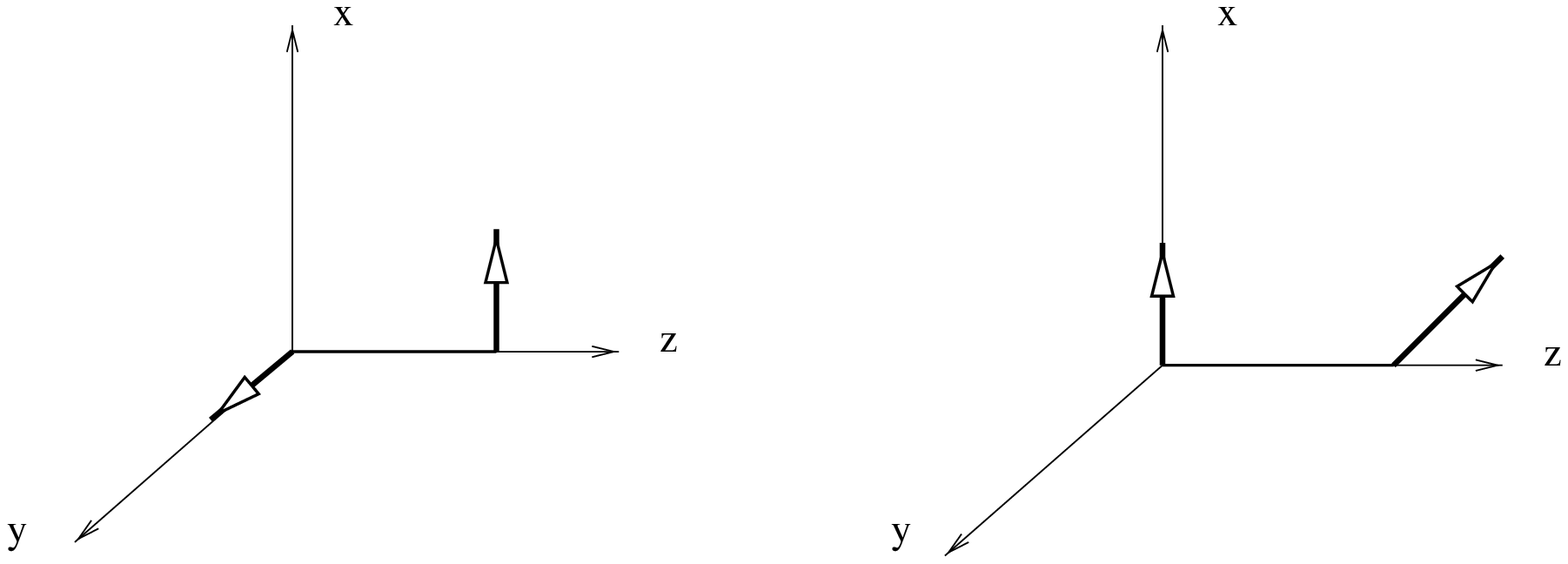}{0.8}
\vskip -10mm
\hskip 40mm {$({\rm {\bf A_1}})$} \hskip 70mm {$({\rm  {\bf A_2}})$}
\vskip -23mm
\hskip 20mm ${\vec b_1}$ \hskip 80mm ${\vec b_1}$
\hskip -55mm ${\vec b_2}$ \hskip 85mm ${\vec b_2}$
\vskip 27mm
\inseps{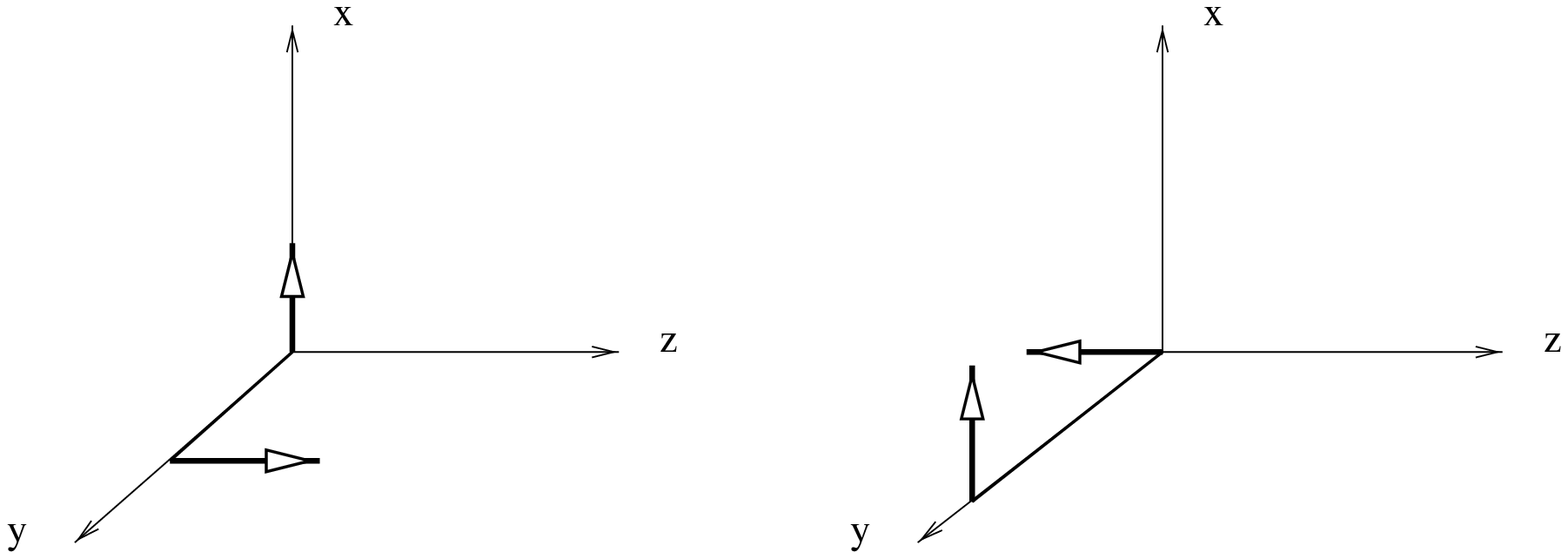}{0.8}

\vskip -10mm
\hskip 40mm {$({\rm {\bf A_3}})$} \hskip 70mm {$({\rm {\bf A_4}})$}
\vskip -23mm
\hskip 23mm ${\vec b_1}$ \hskip 76mm ${\vec b_1}$
\vskip 6mm
\hskip 32mm ${\vec b_2}$ \hskip 49mm ${\vec b_2}$

\vskip 27mm
\inseps{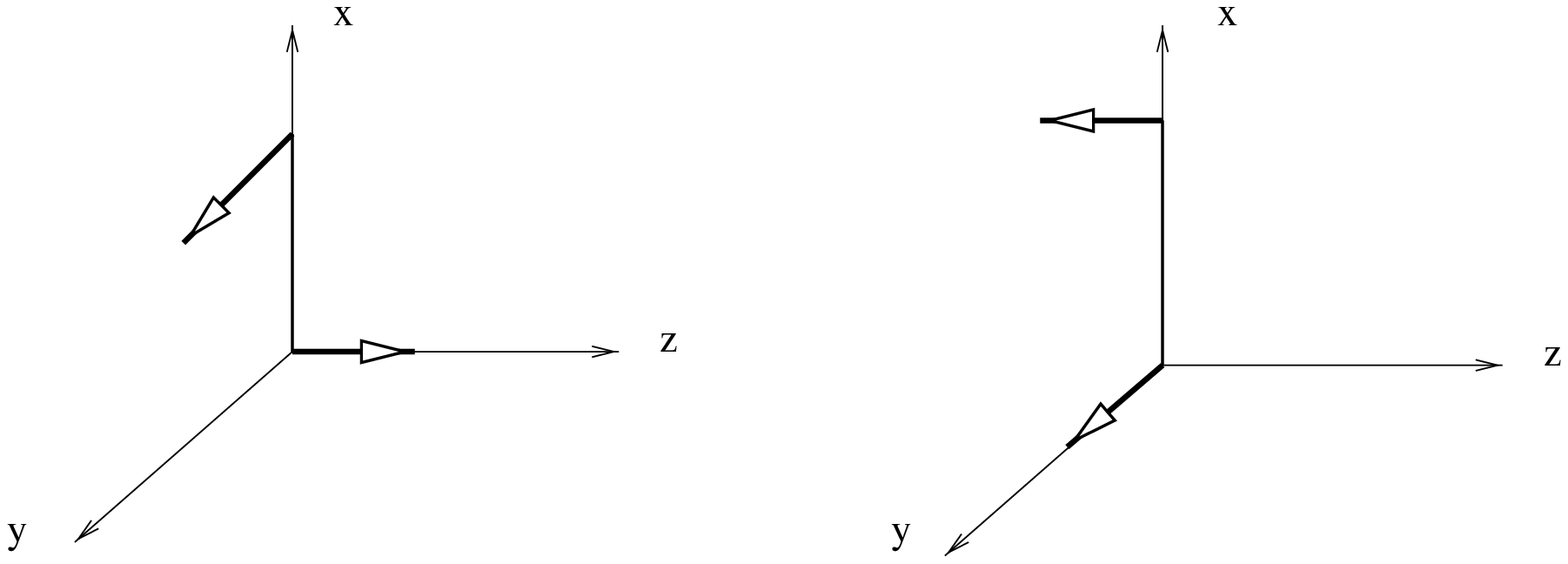}{0.8}
\vskip -10mm
\hskip 40mm {$({\rm {\bf A_5}})$} \hskip 70mm {$({\rm {\bf A_6}})$}
\vskip -45mm
\hskip 25mm ${\vec b_2}$ \hskip 67mm ${\vec b_2}$
\vskip 17mm
\hskip 33mm ${\vec b_1}$ \hskip 66mm ${\vec b_1}$

\vskip 30mm

\end{figure}

\newpage
\centerline{\bf Figure Caption}
\vskip 10mm
{\bf Figure 1}: The allowed orientations (${\rm A_i}$) for a
chiral molecule in the restricted Kuhn model. The displacement of the
electrons are along $\vec b_1$ and $\vec b_2$. We consider a linearly
polarized wave propagating along the $z$ axis, with $\vec {\rm
E}={\rm Re}({\rm E}(z) \ e^{-j\omega t}) \vec e_x$, see text. In orientations
(${\rm A_1}$) and (${\rm A_2}$), where $\vec {b_{12}^{0}}=\vec e_z$,
the electric field is not the same on the two electrons, leading to a
non zero contribution to ${\rm {\vec P}}_{\perp}$, see equation
(19). In orientations (${\rm 
A_3}$) and (${\rm A_4}$), where $\vec {b_{12}^{0}}=\vec e_y$, the 
electric field is the same on the two electrons, leading to a
contribution to ${\rm {\vec P}}_{//}$. In orientations (${\rm A_5}$)
and (${\rm A_6}$), where $\vec {b_{12}^{0}}=\vec e_x$, the electrons
are not coupled to the electric field since ($\vec {\rm E} \cdot \vec
b_1=\vec {\rm E} \cdot \vec b_2 =0$).


\begin{references}
\bibitem{transverse}
As equation (\ref{pol1}) shows, it is consistent to take ${\rm
div} \ {\rm {\vec E}}=0$.
\bibitem{transverse2}
As equation (\ref{pol}) shows, it is again consistent to take ${\rm
div} \ {\rm {\vec E}}=0$.
\bibitem{Thermal}
We neglect here the thermally controlled reactions (${\rm D} \to {\rm
L}$) and (${\rm L} \to {\rm D}$).
\end{references}
\end{document}